\documentclass[aps, prd, twocolumn, groupedaddress ]{revtex4-1}
\usepackage{amsmath,amssymb}
\usepackage{graphicx}
\usepackage{color}
\usepackage{slashed} 
\usepackage{hyperref}
\begin{document}
\newcommand{\nn}{\nonumber}
\def\d{{\mathrm{d}}}
\def\lint{\hbox{\Large $\displaystyle\int$}}   
\def\hint{\hbox{\huge $\displaystyle\int$}}  
\title{\bf\Large Super-radiance and flux conservation}
\author{Petarpa Boonserm\,}
\email[]{petarpa.boonserm@gmail.com}
\affiliation{Department of Mathematics and Computer Science, Faculty of Science, \\
Chulalongkorn University, Bangkok 10330, Thailand}
\author{Tritos Ngampitipan\,}
\email[]{tritos.ngampitipan@gmail.com}
\affiliation{Department of Physics, Faculty of Science,
Chulalongkorn University, Bangkok 10330, Thailand}
\author{Matt Visser\,}
\email[]{matt.visser@msor.vuw.ac.nz}
\affiliation{ \mbox{School of Mathematics, Statistics, and Operations Research,}
Victoria University of Wellington; \\
PO Box 600, Wellington 6140, New Zealand.\\}
\date{28 July 2014; \LaTeX-ed \today}
\begin{abstract}
The theoretical foundations of the phenomenon known as super-radiance still continues to attract considerable attention.  Despite many valiant attempts at pedagogically clear presentations, the effect nevertheless still continues to generate some significant confusion. Part of the confusion arises from the fact that super-radiance in a quantum field theory [QFT] context is not the same as super-radiance (super-fluorescence) in some condensed matter contexts; part of the confusion arises from traditional but sometimes awkward normalization conventions, and part is due to sometimes unnecessary confusion between fluxes and probabilities. We shall argue that the key point underlying the effect is flux conservation, (and, in the presence of dissipation, a controlled amount of flux non-conservation), and that attempting to phrase things in terms of reflection and transmission \emph{probabilities} only works in the \emph{absence} of super-radiance.  To help clarify the situation we present a simple exactly solvable toy model exhibiting both super-radiance and damping. 

\end{abstract}
\maketitle
\def\sign{{\;\mathrm{sign}}}
\def\Dirac{{\slashed D }}
\section{Introduction}

The phenomenon of QFT-induced super-radiance has a long and quite tortuous history. Key high-points are the articles by Zeldovich~\cite{Zeldovich}, and Manogue~\cite{Manogue}, and the more recent work by Richartz \emph{et al}~\cite{Richartz:2009, Richartz:2012}. 
There are close connections with the so-called ``Klein paradox'' for relativistic fermions~\cite{Manogue, Ravndal, Dombey:1, Dombey:2}, and also some significant differences.  
Specific applications to black hole physics include the issues explored in references~\cite{Richartz:2009, Richartz:2012, Press:1972, Winstanley:2001, Cardoso:2004, Dolan:2007, Chen:2008, Kenmoku:2008, Dolan:2012, Herdeiro:2013,  Hod:2013, Degollado:2013}. 
In our own research, when dealing with black hole greybody factors, we have had to deal with super-radiance for Kerr, Kerr--Newman, and Myers--Perry black holes, see~\cite{Kerr-Newman, Myers-Perry} and a related conference article~\cite{Conference}.

Despite all efforts, the super-radiance effect nevertheless still continues to generate significant confusion. Part of the confusion is purely linguistic --- arising from the fact that super-radiance in a traditional QFT context is not the same as super-radiance (super-fluorescence; Dicke super-radiance) in traditional condensed matter contexts~\cite{Dicke}. Part of the confusion arises from the use of utterly traditional and standard but sometimes awkward normalization conventions~\cite{Manogue, Messiah}. Part of the confusion is due to sometimes neglecting the necessary distinction between fluxes and probabilities. 

Extending and modifying the analysis of Richartz \emph{et al}~\cite{Richartz:2009}, we shall argue that the key point underlying the effect is flux conservation, (and, in the presence of dissipation, a controlled amount of flux non-conservation). We shall see that attempting to phrase things in terms of reflection and transmission \emph{probabilities} only works in the \emph{absence} of super-radiance.   

To illustrate and clarify the situation we shall present a particularly simple and exactly solvable toy model, one  which explicitly exhibits both super-radiance and damping. 
While our own interest in these issues was strongly influenced by research into black hole physics, it should be emphasized that the underlying issues and related phenomena are much more general.

\vspace{-10pt}
\section{Super-radiance: Background}
\vspace{-10pt}

One key observation is to note that super-radiance never occurs when one is dealing with the Schr\"odinger equation, and at a minimum requires something like the Klein--Gordon equation~\cite{Kerr-Newman, Myers-Perry}.  For instance, in any axially-symmetric stationary background, once one applies separation of variables $\psi(x,t) = \psi(r,\theta) \; e^{-i\omega t} \; e^{-im\varphi}$ to a neutral scalar field~\cite{Carter, Frolov}, the Klein--Gordon equation becomes
\begin{equation}
\Delta_2 \, \psi(r,\theta) = \left[V(r,\theta) - \left(\omega - m\varpi(r,\theta)\right)^2\right] \; \psi(r,\theta) = 0.\quad
\end{equation}
It is the trailing term in the effective potential, the $\left(\omega - m\varpi\right)^2$ term, that is responsible for the qualitatively new phenomenon of super-radiance, which never occurs in ordinary non-relativistic quantum mechanics.

The reason for this is that the Schr\"odinger equation is first-order in time derivatives, so the effective potential for Schr\"odinger-like barrier-penetration problems is generically of the simple form
\begin{equation}
U(r)   = V(r)  - \omega.
\end{equation}
In contrast, for problems based on the Klein--Gordon equation (second-order in time derivatives) the qualitative structure of the effective potential is
\begin{equation}
U(r) = V(r) - (\omega - m\varpi)^2.
\end{equation}

Similar phenomena occur for charged particles where one has a $(\omega-q\Phi)^2$ contribution to the effective potential.
We shall soon see that it is when the quantity $\omega-m\varpi$, (or more generally, the quantity $\omega-m\varpi-q\Phi$), changes sign that the possibility of super-radiance arises. 
(See for instance the general discussion by Richartz \emph{et al.}~\cite{Richartz:2009, Richartz:2012}.)   
For our purposes in references~\cite{Kerr-Newman, Myers-Perry} super-radiance is related to the rotation of the black hole~\cite{Myers-Perry:1986, Emparan}, but if the scalar field additionally carries electric charge there is a separate route to super-radiance~\cite{Manogue, Ida, Creek1, Creek2}. 

While the Dirac equation, being first-order in both space and time, might seem to completely side-step this phenomenon, it is a standard result that iterating the Dirac differential operator twice produces a Klein--Gordon-like differential equation. In terms of the Dirac matrices we have:
\begin{equation}
\Dirac^2 = 2 (\nabla-iqA)^2 + q F_{ab} \; [\gamma^a, \gamma^b].
\end{equation}
So, once one factors out the spinorial components, and concentrates attention on the second-order differential equation for the amplitude of the Dirac field, even the Klein paradox for charged relativistic fermions can be put into a closely related (though distinct) framework~\cite{Manogue}. It is the trailing $(\omega - m\varpi-q\Phi)^2$ term in the effective potential, and more specifically the change in sign of $\omega - m\varpi-q\Phi$, that is now the harbinger of the so-called ``Klein paradox''. (Which, of course, is not really a ``paradox''~\cite{Manogue, Ravndal, Dombey:1, Dombey:2}.)

\section{Super-radiance: Fluxes}

We shall argue that in the long run it is best to phrase things in terms of relative \emph{fluxes} rather than \emph{probabilities}.  
For a unit incoming flux, consider the equation:
\begin{equation}
F_\mathrm{reflected} + F_\mathrm{transmitted} = 1 -F_\mathrm{dissipated}.
\end{equation}
As long as there is some flux conservation law, as for the Klein--Gordon equation, we can \emph{always} say this, with these signs. 
(Dissipation can be dealt with by giving the potential $V(r,\theta)$ an imaginary contribution, see the discussion below.) 
In \emph{some} cases this general result simplifies, and we can reduce this statement about \emph{fluxes} to a statement about \emph{probabilities}.

\noindent
For example: \\
--- 1) If there is no dissipation ($F_\mathrm{dissipated}=0$), and if the transmitted flux is nonnegative ($F_\mathrm{transmitted} \geq 0$), then we can simply set $R\leftarrow F_\mathrm{reflected}$ and $T\leftarrow F_\mathrm{transmitted}$, and reinterpret these (relative) \emph{fluxes} as \emph{probabilities} with:
\begin{equation}
R + T = 1.
\end{equation}
--- 2) If there is some dissipation ($F_\mathrm{dissipated}>0$), and if the transmitted flux is nonnegative ($F_\mathrm{transmitted} \geq 0$), then we can set $R\leftarrow F_\mathrm{reflected}$ and $T\leftarrow F_\mathrm{transmitted}$  and $P_D\leftarrow F_\mathrm{dissipated}$, and then reinterpret these (relative) \emph{fluxes} as \emph{probabilities} with $P_D$ now being the probability of decay:
\begin{equation}
R + T + P_D = 1.
\end{equation}
--- 3) In contrast, if $F_\mathrm{transmitted} < 0$,  then we \emph{cannot} phrase things in terms of probabilities that add up to $1$. We have to work in terms of fluxes. In particular in this super-radiant regime we have
\begin{equation}
F_\mathrm{transmitted}  = - |t|^2  \leq 0.
\end{equation}
Note the \emph{sign}.
It is the possibility of \emph{negative transmitted flux} that lies at  the heart of super-radiance; in this situation:
\begin{eqnarray}
F_\mathrm{reflected} &=& 1- F_\mathrm{transmitted} - F_\mathrm{dissipated} 
\nonumber
\\
&=& 1 +| F_\mathrm{transmitted}| - F_\mathrm{dissipated}.
\end{eqnarray}
The reflected flux can then easily become over unity.

\section{Super-radiance: Toy model}

To see how this all works in detail, it is best to choose a highly idealized but exactly solvable model.
Working in 1+1 dimensions, consider the PDE
\begin{equation}
\left[ - (\partial_t - i \varpi(x))^2 + c^2\partial_x^2  - V(x) \right] \psi (t,x)=0.
\end{equation}
For simplicity we are working with a massless particle (\emph{eg} photon), as this cuts to the heart of the matter. Adding particle rest masses is not particularly difficult, (see \emph{eg} Manogue~\cite{Manogue}), but adds technical complications that are not central to the issues we wish to discuss.

Taking $\psi(t,x) = e^{-i\omega t} \; \psi(x)$ this is now equivalent to considering the ODE
\begin{equation}
c^2 \; \partial_x^2 \psi(x) = [V(x) - (\omega-\varpi(x))^2]\psi(x).
\end{equation}
Setting $\varpi(x)\to0$ then yields a ``Schr\"odinger-like'' equation, with no possibility of super-radiance, whereas $\varpi(x)\neq 0$ is essential for super-radiance.

Let us now brutally simplify the problem, (in the interests of making it \emph{analytically solvable}), by setting $V(x)\to 0$ and taking
\begin{equation}
\varpi(x) = \Omega \sign(x).
\end{equation}
This toy model is a tractable stand-in for generic situations where $\varpi(x)$ satisfies boundary conditions $\varpi(\pm\infty) =  \pm\Omega$.
We also take units where $c\to1$. Then we are interested in
\begin{equation}
\partial_x^2 \psi(x) =- (\omega-\Omega\sign(x))^2\psi(x).
\end{equation}
We shall soon see that for $|\omega| > |\Omega|$ we obtain ordinary scattering, with no super-radiance;
whereas for  $|\omega| < |\Omega|$ we obtain super-radiance, plus spontaneous emission.

Now for $x\neq0$ this ODE has solutions of the form
\begin{equation}
\psi(t,x) = e^{-i(\omega t- k_\pm x) };
\qquad
k_\pm^2 = (\omega\mp\Omega)^2.
\end{equation}
But which root should we take? As is standard, let us consider the \emph{group velocity}
\begin{equation}
v_g = {\partial\omega\over\partial k_\pm} = {1\over\partial k_\pm/\partial\omega} = {1\over(\omega\mp\Omega)/k_\pm} = {k_\pm\over\omega\mp\Omega}.
\end{equation}
So for the mode with positive group velocity we must have $\sign(k_\pm) = \sign(\omega\mp\Omega)$, whence
\begin{equation}
k_\pm = \sign(\omega\mp\Omega) \; |\omega\mp\Omega| = \omega\mp\Omega; \qquad v_g = +1.
\end{equation}
This is valid for \emph{all} $\omega$, positive or negative. Furthermore
\begin{equation}
k_+k_- = \omega^2-\Omega^2; \quad \sign(k_+k_-) = \sign(\omega^2-\Omega^2).
\end{equation}
Note in contrast that for the phase velocity
\begin{equation}
v_p^\pm = {\omega\over\omega\mp\Omega}.
\end{equation}
This easily flips sign in some regions, in fact
\begin{equation}
\sign(v_p^\pm) = \sign(\omega)\, \sign(\omega\mp\Omega). 
\end{equation}
Now consider something incoming from the left, \emph{and for the time being don't worry about the normalization}.  Matching across the origin we have
\begin{equation}
{e^{ik_- x}}+ r \,{e^{-ik_-x}} \longleftrightarrow t \,{e^{ik_+ x}}.
\end{equation}
From continuity of wavefunction and derivative
\begin{equation}
{1 + r} = {t};  \qquad k_-(1-r) = k_+ \, t.
\end{equation}
Therefore
\begin{equation}
k_-(1-r) = k_+ (1+r),
\end{equation}
implying
\begin{equation}
r = - {k_+-k_-\over k_++k_-} = - {(\omega-\Omega)-(\omega+\Omega)\over(\omega-\Omega)+(\omega+\Omega)} = +{\Omega\over\omega}.
\end{equation}
This is valid for all $\omega$, and normalization independent, (\emph{since the reflected mode automatically has the same normalization as the incoming mode}).  The reflected flux (more precisely, the ratio of reflected to incident flux) is thus 
\begin{equation}
\qquad
F_\mathrm{reflected} = |r|^2 = {\Omega^2\over \omega^2}.
\end{equation}
However, \emph{if we want to fully understand transmitted flux, we need to normalize properly}.

Now consider something incoming from the left,  and normalize relativistically:
\begin{equation}
{e^{ik_- x}\over\sqrt{2|k_-|}}.
\end{equation}
The $\sqrt{2}$ is standard for the relativistic Klein-Gordon equation, to make the flux  simple. One must remember to include both $\psi^*(-i\partial_x) \psi$ \emph{and its hermitian conjugate} when calculating the flux. (For odd historical reasons, for the non-relativistic Schr\"odinger equation people do not put the $\sqrt{2}$ in the normalization of the modes, they instead put an explicit ${1\over2}$ in the definition of the current.)
With this normalization we now have (\emph{note that this new amplitude  ``$t\!$'' will be different from the previous one})
\begin{equation}
{e^{ik_- x} \over\sqrt{2|k_-|}}+ r \,{e^{-ik_-x}\over\sqrt{2|k_-|}} \longleftrightarrow t \,{e^{ik_+ x}\over\sqrt{2|k_+|}}.
\end{equation}
From continuity of wavefunction and derivative we have
\begin{equation}
{1 + r\over\sqrt{2|k_-|}} = {t\over\sqrt{2|k_+|}};  \quad {k_-\over\sqrt{2|k_-|}}\,(1-r) = {k_+\over\sqrt{2|k_+|}} \, t.
\end{equation}
So \emph{we still have}
\begin{equation}
k_-(1-r) = k_+ (1+r),
\end{equation}
implying
\begin{equation}
r = - {k_+-k_-\over k_++k_-} = - {(\omega-\Omega)-(\omega+\Omega)\over(\omega-\Omega)+(\omega+\Omega)} 
= +{\Omega\over\omega}.
\end{equation}
Consequently, as before
\begin{equation}
F_\mathrm{reflected} = |r|^2 = {\Omega^2\over \omega^2}.
\end{equation}
But now, for the transmission amplitude we have
\begin{equation}
t =   \sqrt{|k_+|\over|k_-|} \, \left(1 +{\Omega\over\omega}\right) = \sqrt{|\omega-\Omega|\over|\omega+\Omega|} \left({\omega+\Omega\over\omega}\right).
\end{equation}
--- If $|\omega|>|\Omega|$, (the usual situation), then we see
\begin{eqnarray}
t &=&  \sqrt{\omega-\Omega\over\omega+\Omega} \left[{\omega+\Omega\over\omega}\right] = {\sqrt{\omega^2-\Omega^2}\over \omega} 
\nonumber\\
&=& \sign(\omega)\;  \sqrt{1-{\Omega^2\over\omega^2}},
\end{eqnarray}
and so
\begin{equation}
|t|^2 = 1-{\Omega^2\over\omega^2}\geq 0; \qquad  F_\mathrm{reflected}  +|t|^2= 1.
\end{equation}
So in the usual situation we can meaningfully write
\begin{equation}
F_\mathrm{transmitted} = |t|^2 \geq 0.
\end{equation}

\noindent
--- However, if $|\omega|<|\Omega|$, (the super-radiant case), then
\begin{eqnarray}
t &=&  \sqrt{- {(\omega-\Omega)\over(\omega+\Omega)}} \left({\omega+\Omega\over\omega}\right) 
= {\sqrt{\Omega^2-\omega^2}\over \omega}
\nonumber\\
&=& \sign(\omega)\;  \sqrt{{\Omega^2\over\omega^2}-1},
\end{eqnarray}
and so in this situation
\begin{equation}
|t|^2 = {\Omega^2\over\omega^2}-1;   \qquad F_\mathrm{reflected} - |t|^2 = 1.
\end{equation}
Note the sign flip in the flux conservation law. In the super-radiant situation we \emph{must} write
\begin{equation}
F_\mathrm{transmitted} = - |t|^2 \leq 0.
\end{equation}
To get a deeper understanding of  where the minus sign came from,  note that the flux for a ``properly normalized'' state is 
\begin{equation}
\hbox{(flux)} =   \left( {e^{ik_\pm x}\over\sqrt{2|k_\pm|}} \right)^* \left[ -i\partial_x \left( {e^{ik_\pm x}\over\sqrt{2|k_\pm|}} \right) \right] + \hbox{(conjugate)}.
\end{equation} 
But then
\begin{equation}
\hbox{(flux)}  
= {k_\pm\over|k_\pm|} = \sign(k_\pm) 
= \sign(\omega\mp\Omega).
\end{equation} 
So the flux may not be in the direction one naively expects. 
We can summarize the situation by saying that in \emph{both} cases
\begin{equation}
F_\mathrm{transmitted} = \sign(k_+ k_-) \; |t|^2 = 1 -  {\Omega^2\over\omega^2}.
\end{equation}
This formula is now equally valid for both normal and super-radiant regimes, and for particles incoming from either the left or the right, and easily leads one to verify that in this situation (that is, with no dissipation)
\begin{equation}
F_\mathrm{reflected} + F_\mathrm{transmitted} = 1.
\end{equation}
We could also write this more explicitly as
\begin{equation}
|r|^2 +  \sign(k_+ k_-) \; |t|^2 = 1.
\end{equation}
This is manifestly \emph{not} conservation of \emph{probability}; but is the perhaps more interesting statement that we have conservation of \emph{flux}. In particular, we see that super-radiance can be adequately understood using first-quantization.

\noindent
\emph{Warning:} Because of the way some authors (specifically Manogue~\cite{Manogue}, and Richartz \emph{et al}~\cite{Richartz:2009, Richartz:2012}, and even textbook presentations such as Messiah~\cite{Messiah}), choose to normalize the transmission amplitude, their key result  is instead:
\begin{equation}
|r|^2  + {k_-\over k_+} \; |t|^2 = 1. 
\end{equation}
This is not physically different, but is perhaps a little less transparent. 

\section{Spontaneous emission}

To understand spontaneous emission we need to bring in some foundational ideas from second quantization.
The key point in second quantization is to understand the vacuum state; choosing a vacuum state amounts to (what is called) choosing the division between ``positive and negative frequencies'', an issue which is now just a little more subtle than one might at first expect.
Recall that $k_\pm = \omega \mp \Omega$, and that the unit flux modes are singular at $k_\pm=0$,  (that is at $\omega=\Omega$ in the right-hand half-line, and at $\omega=-\Omega$ in the left-hand half-line). 

\noindent
This observation now leads us, on the two half-lines,  to identify ``particle modes'' as 
\begin{equation}
{\exp(-i[\omega t - [\omega\mp\Omega] x)\over\sqrt{2|\omega\mp\Omega|}};  \qquad \omega > \pm\Omega; \qquad \hbox{(flux)} = +1,
\end{equation}
\begin{equation}
{\exp(-i[\omega t + [\omega\mp\Omega] x)\over\sqrt{2|\omega\mp\Omega|}};   \qquad \omega > \pm\Omega; \qquad \hbox{(flux)}  = -1,
\end{equation}
and to identify ``vacuum modes'' as
\begin{equation}
{\exp(-i[\omega t - [\omega\mp\Omega] x)\over\sqrt{2|\omega\mp\Omega|}};  \qquad \omega < \pm\Omega; \qquad \hbox{(flux)} = -1,
\end{equation}
\begin{equation}
{\exp(-i[\omega t + [\omega\mp\Omega] x)\over\sqrt{2|\omega\mp\Omega|}};   \qquad \omega <\pm\Omega; \qquad \hbox{(flux)} = +1.
\end{equation}
Once these modes have been identified, the rest of the analysis is relatively prosaic.

\noindent
--- For $\omega > |\Omega|$ we are dealing with ``particle modes'' on both sides of the barrier; the usual scattering rules apply, regardless of the direction the particle is initially moving in. 

\noindent
--- For $\omega < -|\Omega|$ we are dealing with ``vacuum modes'' on both sides of the barrier;  this situation is not physically relevant  for our current purposes, regardless of which direction the particle is initially moving in.  

\noindent
--- For $\omega\in(-|\Omega|,+|\Omega|)$, then on one side of the barrier you are dealing with ``particle modes'' and on the other side with ``vacuum modes'', this is the tricky situation. 
Suppose for definiteness $\Omega>0 $ is positive, and  $\omega\in(-\Omega,+\Omega)$, then in the left-hand half-space we are dealing with  particle modes, and in the right-hand half-space we are dealing with vacuum modes.  

For particles incident from the left we have already done the calculation and found super-radiance. In the right hand half space we have a right-moving vacuum mode carrying a leftward flux.
But what happens if a left-moving vacuum mode comes from the right and hits the barrier? It may partially reflect to a right-moving vacuum mode, but partially transmit to form a left moving particle mode in the left hand half space. This is spontaneous emission. Let us do the relevant calculation.
We now have
\begin{eqnarray}
t\; {\exp(-i[\omega t + k_- x])\over\sqrt{2|k_-|}} &\longleftrightarrow& {\exp(-i[\omega t + k_+ x])\over\sqrt{2|k_+|}}  
\nonumber\\
&&+ r\; {\exp(-i[\omega t - k_+ x])\over\sqrt{2|k_+|}}. \qquad
\end{eqnarray}
Continuity of wavefunction and derivatives now implies
\begin{equation}
{t\; \over\sqrt{2|k_-|}} = {1+r \over\sqrt{2|k_+|}}; \qquad {t\;k_- \over\sqrt{2|k_-|}} = {(1-r)k_+ \over\sqrt{2|k_+|}}.
\end{equation}
Note several strategic sign flips compared to the previous calculation.  We now have
\begin{equation}
(1+r)k_- = (1-r) k_+,
\end{equation}
so that
\begin{equation}
 r = {k_+-k_-\over k_+ + k_-}  = {(\omega-\Omega)-(\omega+\Omega)\over(\omega-\Omega)+(\omega+\Omega)} 
 = - {\Omega\over\omega}.
\end{equation}
Similarly
\begin{eqnarray}
t &=& \sqrt{|k_-|\over|k_+|} \left(1- {\Omega\over\omega} \right) = \sqrt{\Omega+\omega\over\Omega-\omega} \left({\omega-\Omega\over\omega} \right) 
\\
&=& {\sqrt{\Omega^2-\omega^2}\over\omega} = \sign(\omega) \sqrt{{\Omega^2\over\omega^2}-1}.
\end{eqnarray}
Since the amplitude $t$ is associated with a left-moving particle in the left half-line,  the flux in the left-hand half-line is
\begin{equation}
\hbox{(flux)} = - |t|^2 = - \left({\Omega^2\over\omega^2}-1\right) < 0. 
\end{equation}
The flux is leftward. Particles are being emitted by the barrier and escaping to the left. (Vacuum modes from the right are escaping from the barrier and moving to the left, the region in which they become particle modes.) Unfortunately this flux is dimensionless, it is a relative flux --- the ratio of the flux of left moving particle modes on the left half-line to the flux of left-moving vacuum modes on the right half-line.

\bigskip
To convert this to an absolute flux we note that the ``unit flux'' condition corresponds to 
\begin{equation}
{\d^2 N\over\d t\; d\omega} = 1.
\end{equation}
That is, one particle \emph{per unit time per unit frequency}. Then the absolute spontaneous emission rate of left-moving particles is
\begin{equation}
{\d^2 N\over\d t\; d\omega} =  \left({\Omega^2\over\omega^2}-1\right);  \qquad \omega^2 \leq \Omega^2.
\end{equation}
Note spontaneous emission occurs only within the super-radiant regime.

\section{Consistency check}

Note that for the specific toy model we have considered, the amplitudes  $t$ and $r$ are infinite at $\omega=0$. An observation along these lines is hidden in Manogue's article~\cite{Manogue}, buried in appendix 1, near the top of page 278. 

Ultimately this infinity is a kinematic singularity due to the fact that $k_+(\omega=0)=- k_-(\omega=0)$.  More generally we could consider a ``shifted'' effective potential by taking
\begin{equation}
\varpi(x) = \bar\Omega + \Delta \sign(x).
\end{equation}
Then whenever one encounters $\pm\Omega$ it would be replaced by $\Omega_\pm = \bar \Omega \pm \Delta$. It is easy to see that one now has
\begin{equation}
k_\pm = \omega\mp \Omega_\pm = (\omega-\bar\Omega)\pm\Delta,
\end{equation}
and that now $k_+k_- = (\omega-\bar\Omega)^2 - \Delta^2$. 
Re-doing the remainder of the relevant calculations one now finds:
\begin{equation}
|r|^2 = {\Delta^2\over(\omega-\bar\Omega)^2}; \qquad |t|^2 = \left| 1 -  {\Delta^2\over(\omega-\bar\Omega)^2}\right|.
\end{equation}
Note one still has
\begin{equation}
|r|^2 +\sign(k_+k_-)\; |t|^2 = 1.
\end{equation}
The kinematic infinity has now moved, from $\omega=0$ to $\omega=\bar\Omega$, but the basic form of the flux conservation law is unaltered. The stability of the flux conservation law under the introduction of and shifts in $\bar\Omega$ is encouraging. 

Indeed, the basic form of the flux conservation law cannot depend on the particular toy model, which was adopted only for simplicity of presentation. 
As long as well defined asymptotic states exist in the infinite left and infinite right, (so $\varpi(\pm\infty)$ must be well-defined and finite), then the form of the relevant second-order ODE guarantees the existence of a transfer matrix~\cite{Transfer, Transfer2},  and also permits (with a suitable change in normalization) a Wronskian analysis along the lines of Richartz \emph{et al}~\cite{Richartz:2009}.

\section{Adding Dissipation}

We had earlier alluded to the fact that dissipation can be modelled by adding an imaginary contribution to the potential.  Let us now see how this works in practice. Set $V(x)\to i \,\Gamma\delta(x)$ so that we are now interested in the ODE
\begin{equation}
\partial_x^2 \psi(x) =[i\,\Gamma \delta(x) - (\omega-\Omega\sign(x))^2]\psi(x).
\end{equation}
For an imaginary delta-function potential the scattering calculation is an easy modification of the quite standard calculation for a real delta-function potential. The key point is that while the wave-function is still continuous at the origin, there will now be a discontinuity in the derivative at the origin:
\begin{equation}
\partial_x \psi(0^+) - \partial_x \psi(0^-) = i \Gamma \; \psi(0). 
\end{equation}

\subsection{Dissipation in Schr\"odinger-like situations}

\noindent
If we (temporarily) set $\Omega\to0$, thereby (temporarily) banishing even the possibility of super-radiance, we will be in a ``Schr\"odinger-like'' situation with damping.
Then matching wave-functions at the origin
\begin{equation}
\exp(+ikx) + r \exp(-ikx) \longleftrightarrow t \exp(+ikx),
\end{equation}
leads to
\begin{equation}
1+r = t; \qquad
\left[  k (1-r) - k t \right] =\Gamma t,
\end{equation}
or equivalently, (since now $k_\pm=k = \omega$ under the current hypotheses), 
\begin{equation}
1+r = t; \qquad
\left[  \omega (1-r) - \omega t \right] =\Gamma t.
\end{equation}
Thence $2  \omega (1-t) = \Gamma t$ and we have
\begin{equation}
t= {\omega\over \omega +{1\over2}\Gamma}.
\end{equation}
Note that $\omega$ is intrinsically positive, and under normal conditions $\Gamma\geq0$. 
The transmission probability is
\begin{equation}
T = |t|^2 = {\omega^2\over (\omega + {1\over2}\Gamma)^2} \in [0,1].
\end{equation}
Now for the reflection amplitude
\begin{equation}
r = t-1 = - {{1\over2}\Gamma\over \omega +{1\over2}\Gamma}.
\end{equation}
Then for the reflection probability
\begin{equation}
R = |r|^2 = {{1\over4}\Gamma^2\over (\omega + {1\over2}\Gamma)^2} \in [0,1].
\end{equation}
But now $T+R\neq 1$ and in fact
\begin{equation}
T+R = 1 - {\omega\Gamma\over (\omega+ {1\over2}\Gamma)^2}.
\end{equation}
So the decay probability is identified as
\begin{equation}
P_D = {\omega\Gamma\over (\omega + {1\over2}\Gamma)^2} \in [0,1].
\end{equation}
This can be viewed as the probability of absorption by the barrier.
Note that 
\begin{equation}
P_D = {\Gamma \; T\over \omega}.
\end{equation}
Dissipation can actually be negative (anti-dissipation) whenever  $\Gamma<0$, (this occurs in non-standard situations where the imaginary part of the potential is negative). 
This observation is compatible with the results of the Wronskian-based analysis of Richartz \emph{et al}~\cite{Richartz:2009}.

\subsection{Dissipation and super-radiance}

Now let us turn $\Omega$ back on, taking $\Omega\neq0$, and see how dissipation interacts with super-radiance, and the mere possibility of having super-radiance. From what we have previously seen, it is now important to focus on \emph{fluxes}, not \emph{probabilities}. 
In first-quantized formalism with the unit flux normalization we wish to match the wavefunctions
\begin{equation}
{e^{ik_- x} \over\sqrt{2|k_-|}}+ r \,{e^{-ik_-x}\over\sqrt{2|k_-|}} \longleftrightarrow t \,{e^{ik_+ x}\over\sqrt{2|k_+|}}.
\end{equation}
From continuity of the wavefunction, and discontinuity of the derivative, we have
\begin{equation}
{1 + r\over\sqrt{2|k_-|}} = {t\over\sqrt{2|k_+|}}; 
\end{equation}
and
\begin{equation}
{k_-\over\sqrt{2|k_-|}}\;(1-r) - {k_+\over\sqrt{2|k_+|}} \; t = {\Gamma\over\sqrt{2|k_+|}} \; t.
\end{equation}
So \emph{we now have}
\begin{equation}
k_-(1-r) - k_+ (1+r) = \Gamma (1+r),
\end{equation}
implying
\begin{equation}
r = - {k_+-k_-+\Gamma\over k_++k_-+\Gamma} 
= - {(\omega-\Omega)-(\omega+\Omega)+\Gamma\over(\omega-\Omega)+(\omega+\Omega)+\Gamma}.
\end{equation}
Consequently,
\begin{equation}
r = {\Omega-{1\over2}\Gamma\over \omega+{1\over2}\Gamma};
\qquad
F_\mathrm{reflected} = |r|^2 = {(\Omega-{1\over2}\Gamma)^2\over (\omega+{1\over2}\Gamma)^2}.
\end{equation}
But now for the transmission amplitude we have
\begin{equation}
t =   \sqrt{|k_+|\over|k_-|} \, \left(1 +{\Omega-{1\over2}\Gamma\over \omega+{1\over2}\Gamma}\right) = 
\sqrt{|\omega-\Omega|\over|\omega+\Omega|} \left({\omega+\Omega\over\omega+{1\over2}\Gamma}\right).
\end{equation}

\bigskip
\noindent
--- If $|\omega|>|\Omega|$, (the non-super-radiant situation), then
\begin{eqnarray}
t &=&  \sqrt{\omega-\Omega\over\omega+\Omega} \left[{\omega+\Omega\over\omega+{1\over2}\Gamma}\right] 
= {\sqrt{\omega^2-\Omega^2}\over \omega+{1\over2}\Gamma},
\end{eqnarray}
and so
\begin{equation}
|t|^2 = {\omega^2-\Omega^2\over(\omega+{1\over2}\Gamma)^2} \geq 0.
\end{equation}
In this non-super-radiant case we can meaningfully write
\begin{equation}
F_\mathrm{transmitted} = |t|^2 = {\omega^2-\Omega^2\over(\omega+{1\over2}\Gamma)^2} \geq 0.
\end{equation}
But now, due to dissipation,  $F_\mathrm{transmitted}+ F_\mathrm{reflected}\neq 1$, and we  in fact have
\begin{eqnarray}
F_\mathrm{dissipated} &=& 1 - F_\mathrm{transmitted} - F_\mathrm{reflected}
\nonumber\\
&=& 1 -  {\omega^2-\Omega^2\over(\omega+{1\over2}\Gamma)^2}  -{(\Omega-{1\over2}\Gamma)^2\over (\omega+{1\over2}\Gamma)^2}
\nonumber\\
&= &
{(\Omega+\omega)\Gamma\over (\omega+{1\over2}\Gamma)^2}.
\end{eqnarray}

\bigskip
\noindent
 --- In contrast, in the super-radiant case, $|\omega|<|\Omega|$, a few key signs flip. We now have
\begin{eqnarray}
t &=&  \sqrt{- {(\omega-\Omega)\over(\omega+\Omega)}} \left({\omega+\Omega\over\omega+{1\over2}\Gamma}\right) 
= {\sqrt{\Omega^2-\omega^2}\over \omega+{1\over2}\Gamma},
\end{eqnarray}
and so in this situation
\begin{equation}
|t|^2 = {\Omega^2-\omega^2\over(\omega+{1\over2}\Gamma)^2} \geq 0.
\end{equation}
In this super-radiant situation we \emph{must} write
\begin{equation}
F_\mathrm{transmitted} = - |t|^2 \leq 0.
\end{equation}

\bigskip
\noindent
--- In either situation, be it super-radiant or normal, we have
\begin{equation}
F_\mathrm{transmitted} = {\omega^2-\Omega^2\over(\omega+{1\over2}\Gamma)^2} = \sign(k_+k_-) \; |t|^2.
\end{equation}
The transmitted flux can be either positive or negative. 
Furthermore, in either situation, be it super-radiant or normal, we now see
\begin{equation}
F_\mathrm{dissipated} = {(\Omega+\omega)\Gamma\over (\omega+{1\over2}\Gamma)^2}.
\end{equation}
Note that 
\begin{equation}
F_\mathrm{dissipated} = {\Gamma \; F_\mathrm{transmitted} \over \omega-\Omega}.
\end{equation}
So again dissipation can actually be negative, (anti-dissipation), if  $\Gamma<0$. (That is, if the imaginary part of the potential is negative). 
This is again compatible with the Wronskian-based analysis of Richartz \emph{et al}~\cite{Richartz:2009}. 

Finally we have
\begin{equation}
F_\mathrm{transmitted}  + F_\mathrm{reflected} + F_\mathrm{dissipated} = 1.
\end{equation}
This formula is now equally valid for both normal and super-radiant regimes, and for particles incoming from either the left or the right.
This is manifestly \emph{not} conservation of \emph{probability}; but is the perhaps more interesting statement that we have conservation of \emph{flux}. In particular, we see that super-radiance can be adequately understood using first-quantization.

\subsection{Dissipation and spontaneous emission}

Spontaneous emission must again be analyzed using some of the foundational ideas from second quantization. 
Fortunately most of the calculation can be easily carried over (with minor modifications) from the dissipation-free case.
Then absolute spontaneous emission rate of particles \emph{per unit time per unit frequency} is:
\begin{equation}
{\d^2 N\over\d t\; d\omega} =  {\Omega^2-\omega^2\over(\omega+{1\over2}\Gamma)^2};  \qquad \omega^2 \leq \Omega^2.
\end{equation}
Note spontaneous emission occurs only within the super-radiant regime.

\section{Discussion}

So in all relevant situations, (without dissipation), with the normalizations of this article we have
\begin{equation}
F_\mathrm{reflected} + F_\mathrm{transmitted} = 1,
\end{equation}
which we can also cast as
\begin{equation}
|r|^2 +  \sign(k_+ k_-) \; |t|^2 = 1.
\end{equation}
This is a very clean and convincing result, which clearly summarizes many of the most important situations.
In the presence of dissipation we must instead write
\begin{equation}
F_\mathrm{reflected} + F_\mathrm{transmitted} = 1 - F_\mathrm{dissipated}.
\end{equation}

For our particular toy model 
\begin{equation}
\partial_x^2 \psi(x) =[i\Gamma \delta(x) - (\omega-\Omega\sign(x))^2]\psi(x),
\end{equation}
we were able to explicitly evaluate
\begin{equation}
F_\mathrm{reflected} = {(\Omega-{1\over2}\Gamma)^2\over (\omega+{1\over2}\Gamma)^2};
\quad
F_\mathrm{transmitted} = {\omega^2-\Omega^2\over(\omega+{1\over2}\Gamma)^2};
\end{equation}
and
\begin{equation}
F_\mathrm{dissipated} = {(\Omega+\omega)\Gamma\over (\omega+{1\over2}\Gamma)^2}.
\end{equation}
If the last two quantities are non-negative, (the first is automatically so), then these fluxes can be reinterpreted in terms of probabilities: 
$R$, $T$, and $P_D$, for reflection, transmission, and decay, respectively.  That is
\begin{equation}
R + T + P_D = 1.
\end{equation}
However, if either of the last two quantities is negative, (either due to super-radiance or anti-damping), then  the formulation in terms of fluxes is more fundamental, and discussion of probabilities should be completely avoided.

\section*{Acknowledgments}

This research has been supported by a grant for the professional development of new academic staff from the Ratchadapisek Somphot Fund at Chulalongkorn University, by the Thailand Toray Science Foundation (TTSF), by the Thailand Research Fund (TRF), by the Office of the Higher Education Commission (OHEC), Chulalongkorn University, and by the Research Strategic plan program (A1B1), Faculty of Science, Chulalongkorn University (MRG5680171). 
PB was additionally supported by a scholarship from the Royal Government of Thailand. 
TN was also supported by a scholarship from the Development and Promotion of Science and Technology talent project (DPST). 
MV was supported by the Marsden Fund, and by a James Cook fellowship, both administered by the Royal Society of New Zealand.




\end{document}